\documentclass[10pt,journal,compsoc]{IEEEtran}

\usepackage{tabularx}
    \newcolumntype{L}{>{\raggedright\arraybackslash}X}
\usepackage{comment}
\usepackage{amsmath}
\usepackage{amssymb}
\usepackage{gensymb}
\usepackage{graphicx}
\usepackage{epstopdf} % Place after graphicx package.
\usepackage{color}
\usepackage{enumerate}
\usepackage{multirow}
\usepackage{cite}
\usepackage{multicol}
\usepackage{threeparttable}
\usepackage[ruled]{algorithm}
\usepackage[noend]{algorithmic}
\usepackage{subfig}
\usepackage{float}
\usepackage{mathtools}
\usepackage{array}
\newcolumntype{P}[1]{>{\centering\arraybackslash}p{#1}}
\newcolumntype{M}[1]{>{\centering\arraybackslash}m{#1}}
\usepackage{nopageno}

\usepackage[numbers]{natbib} 

\begin{document}
\title{Secret Sharing for Generic Theoretic Cryptography}
\author{James Smith}
\date{}
\maketitle
\begin{abstract} 
Sharing a secret efficiently amongst a group of participants is not easy since there is always an adversary / eavesdropper trying to retrieve the secret. In secret sharing schemes, every participant is given a unique share. When the desired group of participants come together and provide their shares, the secret is obtained. For other combinations of shares, a garbage value is returned. A threshold secret sharing scheme was proposed by Shamir and Blakeley independently. In this (n,t) threshold secret sharing scheme, the secret can be obtained when at least $t$ out of $n$ participants contribute their shares. This paper proposes a novel algorithm to reveal the secret only to the subsets of participants belonging to the access structure. This scheme implements totally generalized ideal secret sharing. Unlike threshold secret sharing schemes, this scheme reveals the secret only to the authorized sets of participants, not any arbitrary set of users with cardinality more than or equal to $t$. Since any access structure can be realized with this scheme, this scheme can be exploited to implement various access priorities and access control mechanisms. A major advantage of this scheme over the existing ones is that the shares being distributed to the participants is totally independent of the secret being shared. Hence, no restrictions are imposed on the scheme and it finds a wider use in real world applications. 
\end{abstract}

\section{Introduction}
Secret sharing schemes are important premises in multi-party computation schemes~\cite{wang2015brief}, cloud computing~\cite{singh2017cloud} and cyberphysical systems~\cite{saha2022system, brown2021gravitas}. In a secret sharing scheme a dealer has a secret key. There is a finite set $P$ of participants and a set $\gamma$ of subsets of $P$. A secret sharing scheme with $\gamma$ as the access structure is a method which the dealer can use to distribute shares to each participant so that a subset of participants can determine the key if and only if that subset is in $\gamma$. The share of a participant is the information sent by the dealer in private to the participant. A secret sharing scheme is ideal if any subset of participants who can use their shares to determine any information about the key can in fact actually determine the key, if the set of possible shares is the same as the set of possible keys. \par
The threshold secret sharing schemes proposed in~\cite{DBLP:journals/cacm/Shamir79} and~\cite{DBLP:conf/eccsp/BlakleyK93} are now being used widely in numerous real world applications~\cite{saha2021sharks, saha2022sharks, saha2022machine}. Some of the applications which require the frequent use of secret sharing schemes are cloud computing, banking sectors and sensor networks~\cite{DBLP:conf/sp/ChanPS03}. For the past ten years, threshold implementations are also used for masking to combat side channel attacks~\cite{DBLP:conf/icics/NikovaRR06, saha2016tv, sehwag2016tv}. Unfortunately, with the advent of cloud computing, internet of things and big data, many desired applications cannot be implemented by just threshold implementations. These require more generalized secret sharing schemes which can implement any access structure with more restrictions. Many works on generalized secret sharing schemes exist in the literature but most of these are infeasible in the practical scenario. This is due to the requirement of an enormous number of shares for every user. As proposed in~\cite{ito1989secret}, monotone access structures can be theoretically implemented but in the worst case, each of the $n$ users has to hold on to $2^n$ shares. In such a scenario, the key management system becomes too convoluted and practically unrealizable.\par
It is proven in~\cite{DBLP:conf/crypto/Leichter88} that there exists no threshold secret sharing scheme for arbitrary monotone functions. It has been shown that a fully generalized secret sharing scheme for any arbitrary $\gamma$ does not exist without $2^n$ space and time complexity. In~\cite{DBLP:conf/eccsp/BlakleyK93}, multi-level and compartmented secret sharing schemes are implemented which are more general than threshold secret sharing schemes. Secret sharing schemes have also been generated from lattices~\cite{yamamoto1998proposal, saha2022machine5g}. Each of these schemes are suitable for specific applications but none of them is fully generalized such that any arbitrary $\gamma$ can be implemented. \par   
In this paper, we propose an algorithm to realize fully generalized secret sharing scheme for arbitrary access structures. This implies that a group of participants, say $\alpha$, can access the secret if and only if $\alpha$ $\epsilon$ $\gamma$. All other groups of participants, including all subsets and supersets of $\alpha$, will get a garbage value if they input their shares in the algorithm. If $\gamma$ is a monotone access structure, then all the supersets of the authorized sets present in $\gamma$ will also be able to gain access.\par
The paper first outlines the famous threshold secret sharing scheme. Then we propose our generalized secret sharing scheme followed by its security proof. We describe a possible attacker model and how to prevent the adversary from benefiting from it. The next section outlines the secure parameters to be used for the implementation of the algorithm followed by a section stating some of the possible future works related to this scheme. In the end, we conclude the paper.
\section{Secret Sharing}
Let $S$ be a finite domain of secrets. A secret sharing scheme realizing an access structure $\gamma$ is a scheme in which the input of the dealer is a secret $s$ $\epsilon$ $S$ such that the following two requirements hold:
\begin{itemize}
\item \textbf{Reconstruction requirement:} The secret $s$ can be reconstructed by any authorized set. That is, for any set $G$ $\epsilon$ $\gamma$ $(G=\{i_1,...,i_{|G|}\})$, there exists a reconstruction function $h_G : S_{i_1} \times ... \times S_{i_{|G|}} \Rightarrow S$ such that for every secret $s$ and every random input $r$,
\vspace*{0.3\baselineskip}
\begin{center}
if $\pi(s,r)$ = $\langle s_1,s_2,...,s_n \rangle$ then $h_G\{s_{i_1},...,s_{i_{|G|}\}}$ = $s$
\end{center}
\vspace*{0.2\baselineskip}
\item \textbf{Security Requirement:} Every unauthorized set of participants cannot reveal any partial information about the secret $s$. Stating this explicitly: for any $B$ $\notin$ $\gamma$, for every two secrets $a_1,a_2$ $\epsilon$ $S$, and for every vector of possible pieces $\{s_i\}_{i \epsilon B}$ :
\vspace*{0.2\baselineskip}
\begin{center}
Pr [$\Lambda_{P_i \epsilon B}$ $\pi_i(a_i,r)$ = $s_i$] = Pr [$\Lambda_{P_i \epsilon B}$ $\pi_i(a_2,r)$ = $s_i$]
\end{center}
\vspace*{0.2\baselineskip}
where the probabilities are taken over the random input of the dealer.
\end{itemize}
\section{Threshold Secret Sharing Scheme}

The $(n,t)$ threshold secret sharing scheme states that the secret $S$ is divided into $n$ shares and each of the $n$ participants receive one share each.
\begin{itemize}
\item The secret $S$ can be reconstructed if at least $t$ participants come together and use their shares.
\item The secret $S$ cannot be reconstructed if less than $t$ participants come together and combine their shares.
\end {itemize}

A simple model is discussed over here for the purpose of introduction of the scheme. A polynomial, 
\begin{center}
$y(x) = a_{t-1}x^{t-1}+a_{t-2}x^{t-2}+....+a_{1}x+S$, $a_{i}$  $\epsilon$ $\mathbb{R}$
\end{center}
 is constructed where $S$ is the secret. The shares of the secret given to each user are the points $(x_i,y_i)$ which lie on the $y(x) =0$ curve in the real plane. To reconstruct the secret from the shares, the coefficients of the polynomial have to be calculated. This can be done by solving the system of linear equations and obtaining $S$. The system of linear equations can be obtained by substituting at least $t$ pairs of $(x_i,y_i)$ in the above polynomial and assigning $y(x)=0$. At least $t$ such shares are required to do this as this is a polynomial of degree $(t-1)$ with $t$ unknown coefficients. If less than $t$ shares are available, then the system of equations for solving for $a_i$, $i$  $\epsilon$ $\{1,2,...,(t-1)\}$ and $S$ will have infinite solutions. Hence, this is an efficient implementation of the $(n,t)$ threshold secret sharing scheme. Algorithms for solving a system of linear equations using interpolation techniques show that this computation can be done in $O(nlog^2 n)$ complexity.
 
 \section{Generalized Secret Sharing Scheme}
 In this section, we discuss the proposed generalized secret sharing scheme. First, we discuss how the dealer distributes the shares among the participants followed by the method to retrieve the secret. The next part of this section outlines the security proof of this scheme. The last part demonstrates the implementation of monotone access structures using this scheme.
 
 \subsection{Distributing the shares}
 Let the participants be $\{A,B,C,D,E,F,G\}$ and $\gamma$ be the access structure. The steps for the generalized secret sharing scheme are stated below: \par
 \begin{enumerate}
 \item \textit{Share Distribution Mechanism :} Every user is given a unique prime number as a share. In the aforementioned example, the primes distributed to $A,B,...,G$ are $p_A,p_B,...,p_G$ respectively.
 \vspace*{0.5\baselineskip}
 \item \textit{Characterizing each subset in $\gamma$ :} Every subset $i$ belonging to $\gamma$ is assigned a characteristic number $c_i$ such that,
  \begin{center}
  $c_i$ = ${\displaystyle \prod_{\forall j \epsilon i} p_j}$
  \end{center}
  As every subset $i$ consists of a unique set of participants, every $c_i$ is a product of a unique set of prime numbers. This ensures that $c_i$ is exclusive to $i$.
 \vspace*{0.5\baselineskip}
 \item \textit{Polynomial Construction:} When a subset $i$ of participants contribute their shares, they retrieve the secret $S$ if $i$  $\epsilon$ $\gamma$. Otherwise a garbage value is obtained. To execute this scheme, we construct the following polynomial $y(x)$ Let $k$ be the number of subsets of participants present in $\gamma$.
 \vspace*{0.2\baselineskip}
 \begin{center}
  $y(x)$ = $(x-c_1)(x-c_2)...(x-c_k)+S$
 \end{center}
  \vspace*{0.2\baselineskip}
  Unlike threshold secret sharing schemes, this polynomial is expanded and made public to all the participants.
 \end{enumerate}
 \subsection{Retrieving the secret}
 The steps to retrieve the secret are described below:
  \vspace*{0.3\baselineskip}
 \begin{enumerate}
 \item The set of participants (let it be denoted by $\alpha$) contribute their individual shares and compute $r$.
 \begin{center}
 $r$ = ${\displaystyle \prod_{i=1}^{n} p_{\alpha_i}}$
 \end{center}
 where $n$ is the number of participants in $\alpha$, $\alpha_i$ denotes the $i^{th}$ participant in $\alpha$ and $p_{\alpha_i}$ denotes the prime number assigned to the participant $\alpha_i$.
 \vspace*{0.5\baselineskip}
 \item $y(r)$ is computed by the participants. Every participant had received a polynomial $y(x)$. $y(r)$ can be easily computed by putting $x=r$ in the polynomial.\par
 If $\alpha$ $\epsilon$ $\gamma$, $y(r)$ = $S$. Otherwise $y(r)$ returns a garbage value. 
 \end{enumerate}
 \subsection{Implementing Monotone Access Structures}
 \vspace*{0.25\baselineskip}
 \textit{Definition:} Let $\{P_1,P_2,...P_n\}$ be the set of participants. A collection $\gamma$ $\subseteq$ $2^{\{P_1,...,P_n\}}$ is monotone if $B$ $\epsilon$ $\gamma$ and $B$ $\subseteq$ $C$ implies $C$ $\epsilon$ $A$. \par
 Our scheme does not implement monotone access structures implicitly. Only the sets of users present in $\gamma$ are treated as authorized sets of users. Since this scheme is highly generic, monotone access structures can also be generated using this scheme. To realize a monotone access structure $A$, we create a new access structure $A'$ such that all the possible supersets of the authorized sets present in $A$ are included in $A'$. Formally,
 \begin{center}
 $A'$ = $A$ $U$ $D$\\
 \textit{where}  $D$ = $\{\alpha '$ $|$ $\alpha \subset \alpha '$, $\alpha$ $\epsilon$ $A\}$
 \end{center}
 Then, we design our aforementioned secret sharing scheme for the access structure $A'$.\par
 
 The performance of this algorithm deteriorates significantly when it is modified to implement monotone functions. The time required for this algorithm is of the order $O(k)$, where $k$ is the degree of the polynomial $y(x)$.
 \begin{align}
 k &= \textit{degree of y(x)}\\
     &= n(A)
 \end{align}
 When the access structure $A$ is modified to $A'$, the cardinality increases exponentially. Let the number of authorized sets in $A$ be $k$, the average number of participants in an authorized set be $v$ and the total number of participants be $n$. 
 \begin{center}
 If   $n(A)$ = $k \times v$\\
 then   $n(A')$ = $O(k.2^{n-k})$ 
 \end{center}  
 \section{Proof of Security}
 A secret sharing scheme is said to be secure if it fulfils the following comditions:
 \begin{itemize}
 \item A subset of participants (let it be denoted by $\alpha$) should be able to retrieve the secret deterministically if and only if $\alpha$ $\epsilon$ $\gamma$.
 \item Any other set of participants should not be able to retrieve the secret deterministically, that is they can retrieve the secret with a negligible probability. 
 \item Proper subsets of $\alpha$ should not be able to retrieve the secret.
 \item If the secret sharing scheme implements monotone access structures, then a subset of participants $\beta$, where $(\beta$ $\neq$ $\alpha)$, can retrieve the secret if $\alpha$  $\subseteq$ $\beta$. \par
 \end{itemize}
  \vspace*{1\baselineskip}
  The security proof for this scheme is outlined below. Prior to that we show that,
  \begin{gather}
     y(x) = (x-c_1)(x-c_2)...(x-c_k)+S\\
     \Rightarrow y(x) = g(x)+S\\[10pt]
  \nonumber where, \hspace{10pt} g(x)=(x-c_1)(x-c_2)...(x-c_k)
  \end{gather}
  Let the set of participants contributing their shares be denoted by $\beta$.
  Let
  \begin{center}
  $\gamma$ = $\{\alpha_1,\alpha_2,...,\alpha_k\}$
  \end{center}
  where $\alpha_i$ ($i$ $\epsilon$ $\{1,2,...,k\}$) is a set of participants allowed to access the secret.
  \vspace*{1\baselineskip}
  \begin{itemize}
  \item \textbf{$\beta$ = $\alpha_i$} : This is the case when a set of $n$ participants belonging to $\gamma$ come together to retrieve the secret. The value $r$ is computed as per the method of retrieval of the secret.
  \begin{center}
   $r$ = ${\displaystyle \prod_{j=1}^{n} p_{\beta_j}}$\\
   \vspace*{0.5\baselineskip}
   $\Rightarrow$ $r$ = ${\displaystyle \prod_{j=1}^{n} p_{\alpha_{i,j}}}$\\
   \vspace*{1\baselineskip}
   $\Rightarrow$ $r$ = $c_i$
   \end{center}
   \vspace*{0.7\baselineskip}
   $c_i$ is a root of the polynomial $g(x)$.
   \vspace*{0.7\baselineskip}  Therefore,
   \begin{center}
   $g(r)$ = $g(c_i)$ = $0$\\
   \vspace*{0.5\baselineskip}
   $y(r)$ = $g(r)$ + $S$\\
   $\Rightarrow$ $y(r)$ = $S$
   \end{center}
   \vspace*{1\baselineskip}
   Hence, the secret $S$ is successfully retrieved.
   
\vspace*{1\baselineskip}
\item \textbf{$\beta$ $\neq$ $\alpha_i$} : This is the case when a set of $n$ participants not belonging to $\gamma$ come together to retrieve the secret. The value $r$ is computed as per the protocol.
  \begin{center}
   $r$ = ${\displaystyle \prod_{j=1}^{n} p_{\beta_j}}$\\
   \vspace*{0.5\baselineskip}
   $r$ $\neq$ ${\displaystyle \prod_{j=1}^{n} p_{\alpha_{i,j}}}$\\
   \vspace*{1\baselineskip}
   $\Rightarrow$ $r$ $\neq$ $c_i$\\
   \vspace*{0.5\baselineskip}
   $\Rightarrow$ $g(r)$ $\neq$ $0$\\
   \vspace*{0.5\baselineskip}
   $y(r)$ = $g(r)$ + $S$\\
   $\Rightarrow$ $y(r)$ $\neq$ $S$
   \end{center}
   \vspace*{1\baselineskip}
   Hence, the secret $S$ is not retrieved. As every participant has a unique prime number, it is ensured that $c_i$ can be constructed only when the participants of $\alpha_i$ contribute their shares. This property prevents any other set of participants from retrieving the secret.
  \end{itemize}
 \vspace*{1\baselineskip}
 It is not possible for any adversary to deterministically decompose $y(x)$ into $g(x)+S$. This is because there are $(k+1)$ variables, namely $c_1,c_2,...,c_k$ and $S$. The adversary has access to only $k$ equations. These equations are obtained by equating the algebraic coefficients of $x^{(k-1)},x^{(k-2)},...,x$ and the constant of the polynomial $y(x)$ to their corresponding numerical values.
 \section{An Attacker Model}
 We show that an attacker can guess a range of values in which $S$ lies. However, we can reinforce the security of the scheme by maximizing this range so that the adversary cannot guess the secret effectively.\par
 The attacker knows that $y(x)=g(x)+S$ and $g(x)$ has $k$ roots. He computes the values $\delta_1$ and $\delta_2$ such that
 \begin{center}
 $z_1(x)$ = $y(x)$ + $\delta_1$\\
 $z_2(x)$ = $y(x)$ + $\delta_2$\\
 \end{center}
 so that $\delta_1$ and $\delta_2$ are the minimum and maximum values respectively for which $z_1(x)$ and $z_2(x)$ have exactly $k$ roots. The secret $S$ $\epsilon$ $(-\delta_2,-\delta_1)$.
 
 \begin{figure}[h]
 \centering
 \includegraphics[scale=0.3]{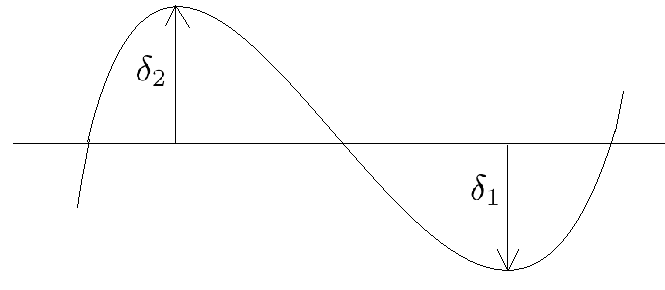} 
 \caption{The values of $\delta_1$ and $\delta_2$ for $y(x) = a_3x^3+a_2x^2+a_1x+a_0$} 
 \label{fig:Attack}
 \end{figure}
 
 To make this attacker model inefficient for the adversary, the function $y(x)$ should be modified appropriately to maximize the values of $\delta_1$ and $\delta_2$.
 
 \section{Parameters of Implementation}
 \vspace*{0.3\baselineskip} 
The polynomial $y(x)$ cannot be computationally decomposed to $g(x) + S$ deterministically. Still, to ensure maximum security of this scheme we choose large prime numbers to be distributed among the participants. Prime factorization being a well-known hard problem, makes the decomposition of $y(x)$ into $g(x)+S$ virtually impossible. The security can be further increased by choosing the primes to be of the same order so that no information of any kind is leaked to the adversary. \par
To ensure the efficient performance of the scheme, it is desirable to restrict the number of authorized sets from becoming too high. The time required to compute $y(x)$ for a given value of $x$ is $O(k)$, where $k$ is the degree of $y(x)$. The space complexity for storing all the coefficients of $y(x)$ is $\theta(k)$. As shown earlier, 
\begin{center}
$k$ = $n(\gamma)$
\end{center}
The scheme becomes computationally infeasible if the $k=O(2^n)$. This is not a problem in real world applications where $k<<2^n$ but for optimal theoretical security and efficiency we would restrict $k$ to be at most a subexponential function of $n$.
 
 \section{Future Work}
 Future work on this scheme may include making it more robust and effective when the number of authorized sets increase. Scope of research lies in finding new adversary models and measures to prevent them. Security bounds can formally be defined by applying intricate complexity theory on this scheme. Optimizations can be done on this scheme to make it more effective for specific applications.
 
 \section{Conclusion}
 This novel scheme is one of the first schemes to realize (strict) generalized ideal secret sharing. The previous schemes have restricted themselves to the realization of monotone access structures. This secret sharing scheme can be utilized effectively in implementing access control and authentication of groups in an environment with a large number of participants. This includes applications in cloud computing, banking sectors and internet of things. This scheme is lightweight, hence it can also be incorporated into resource constrained environments such as embedded devices.

\bibliographystyle{IEEEtranN}
\bibliography{main}
\end{document}